\documentclass[sigconf]{acmart}

\usepackage{flushend}
\usepackage{subfiles}
\usepackage{enumitem}
\usepackage{color, colortbl}
\usepackage{tikz}
\usepackage{caption}
\usepackage{comment}
\usepackage{tabularx, color, colortbl}
\usepackage{hyperref}       % hyperlinks
\usepackage{url}            % simple URL typesetting
\usepackage{booktabs}       % professional-quality tables
\usepackage{amsfonts}       % blackboard math symbols
\usepackage{nicefrac}       % compact symbols for 1/2, etc.
\usepackage{microtype}      % microtypography
\usepackage{lipsum}
\usepackage{amsthm}
\usepackage[normalem]{ulem}
\usepackage[vlined, ruled, linesnumbered]{algorithm2e}
\usepackage{algpseudocode}
\usepackage{subcaption}
\usepackage{siunitx}        % to use round-precision
\renewrobustcmd{\bfseries}{\fontseries{b}\selectfont}
\renewrobustcmd{\boldmath}{}
\newrobustcmd{\B}{\bfseries}
\definecolor{Gray}{gray}{0.9}

\usepackage{pifont}   % for tick and cross marks
\usepackage{float}    % for table placement
\usepackage{xcolor}   % for colored text
\usepackage{cleveref}
\crefname{section}{§}{§§}
\Crefname{section}{§}{§§}
\usepackage[most]{tcolorbox}

\setcopyright{none}
\renewcommand\footnotetextcopyrightpermission[1]{}

\author{Chris Egersdoerfer}
\affiliation{
  \institution{University of Delaware}
  \city{Newark}
  \country{United States}
}
\email{cegersdo@udel.edu}

\author{Philip Carns}
\affiliation{
  \institution{Argonne National Laboratory}
  \city{Lemont}
  \country{United States}
}
\email{carns@mcs.anl.gov}

\author{Shane Snyder}
\affiliation{
  \institution{Argonne National Laboratory}
  \city{Lemont}
  \country{United States}
}
\email{ssnyder@mcs.anl.gov}

\author{Robert Ross}
\affiliation{
  \institution{Argonne National Laboratory}
  \city{Lemont}
  \country{United States}
}
\email{rross@anl.gov}

\author{Dong Dai}
\affiliation{
  \institution{University of Delaware}
  \city{Newark}
  \country{United States}
}
\email{dai@udel.edu}

\begin{document}

%\title{STELLAR: Autonomous LLM Tuning Agent for HPC Storage}
\title{STELLAR: Storage Tuning Engine Leveraging LLM Autonomous Reasoning for High Performance Parallel File Systems}

\begin{abstract} 
I/O performance is crucial to efficiency in data-intensive scientific computing; but tuning large-scale storage systems is complex, costly, and notoriously manpower-intensive, making it inaccessible for most domain scientists.
To address this problem, we propose \texttt{STELLAR}, an autonomous tuner for high-performance parallel file systems. 
Our evaluations show that \texttt{STELLAR} almost always selects near-optimal configurations for the parallel file systems within the first \textit{five attempts}, even for previously unseen applications.

\texttt{STELLAR}’s human-like efficiency is fundamentally different from existing autotuning methods, which often require hundreds of thousands of iterations to converge. \texttt{STELLAR} achieves this through autonomous end-to-end agentic tuning. Powered by large language models (LLMs), \texttt{STELLAR} is capable of (1) accurately extracting tunable parameters from software manuals, (2) analyzing I/O trace logs generated by applications, (3) selecting initial tuning strategies, (4) rerunning applications on real systems and collecting I/O performance feedback, (5) adjusting tuning strategies and repeating the tuning cycle, and (6) reflecting on and summarizing tuning experiences into reusable knowledge for future optimizations. 

\texttt{STELLAR} integrates retrieval-augmented generation (RAG), external tool execution, LLM-based reasoning, and a multiagent design to stabilize reasoning and combat hallucinations. 
We evaluate how each of these components impacts optimization outcomes, thus providing insight into the design of similar systems for other optimization problems.
\texttt{STELLAR}'s architecture and empirical validation open new avenues for tackling complex system optimization challenges, especially those characterized by vast search spaces and high exploration costs. Its highly efficient autonomous tuning will broaden access to I/O performance optimizations for domain scientists with minimal additional resource investment.
\end{abstract}

\maketitle % should come after the abstract

\begingroup
\renewcommand\thefootnote{}
\footnotetext{Preprint. Original copy published in the Proceedings of the 2025 International Conference for High Performance Computing, Networking, Storage, and Analysis (SC25).}
\endgroup

\pagestyle{plain} % should come right after \maketitle
  
\section{Introduction}
\label{sec:intro}
%Today, the large amount of data-intensive scientific applications have turned parallel file system (PFS) a critical component of high-performance computing (HPC). \textcolor{blue}{Some details or numbers about data-intensive applications.}

To deliver high performance, modern parallel file systems (PFSs) expose hundreds of configurable knobs to control I/O behaviors. {For example, Lustre 2.12.5 has at least 159 tunable user parameters~\cite{lustre_manual}. The newest Ceph Nautilus comes with 1,536 parameters, although some of them might not be tunable~\cite{lyu2020sapphire}.} 
Determining the best I/O configurations to achieve optimized I/O performance for each individual application is a crucial task in high-performance computing (HPC) storage. Manual configuration is often impractical, however, because of the vast configuration space, the complexity of underlying hardware, and the diversity of scientific application workloads~\cite{dorier2022hpc}. 
In practice, HPC system administrators will conduct benchmark runs during the installation phases and provide recommended configurations for the entire system~\cite{nersciotuning2025}. This manpower-intensive process is costly and slow. More importantly, the resulting recommendations cannot capture the nuances of all workloads and will be suboptimal for some applications and I/O patterns.  

Recently, automatic tuning methods that leverage heuristic rules ~\cite{li2015ascar,kim2019dca,neuwirth2017automatic}, machine learning ~\cite{cao2018towards,lyu2020sapphire,behzad2019optimizing}, and reinforcement learning~\cite{cheng2021aioc2,li2017capes,zhu2022magpie} have shown great potential. {For instance, ASCAR\cite{li2015ascar} proposed rule-based heuristics to respond to burst I/Os, SAPPHIRE~\cite{lyu2020sapphire} utilized Bayesian optimization to recommend the best configurations, and CAPES~\cite{li2017capes} leveraged reinforcement learning to tune parallel file system parameters}. Although these methods are able to find configurations that outperform system defaults, they all incur significant tuning costs~\cite{cao2018towards}. It has been repeatedly shown that the state-of-the-art autotuning frameworks require hundreds to thousands of iterations or training samples to explore the vast search space, which is formulated by the large number of tunable parameters and large number of choices for each parameter. Such high exploration costs make it impractical for existing autotuning frameworks to tune I/O performance for real-world applications in production environments.

In light of this situation, we draw inspiration from observing how human experts tune parallel file systems.
When faced with a new application, HPC I/O experts typically begin by examining its I/O trace logs (e.g., Darshan logs) to understand the application's defining I/O patterns. Once the I/O patterns are understood, human experts rely on their knowledge to determine an initial tuning strategy. This knowledge was derived from the file system manuals, hardware specifications of the current cluster, and, most importantly the experts’ prior experiences working with other applications on the same HPC platform. Using this knowledge, human experts can identify the most critical parameters to tune, the appropriate value ranges for each parameter, and the tuning direction. For example, for applications with large files shared across many processes, random I/O operations may benefit from larger stripe sizes and stripe counts~\cite{nersciotuning2025}.

After deciding on an initial optimized configuration, human experts typically run the application to validate their tuning strategy. In many cases, a full-scale run is conducted to accurately assess the application’s performance in a production-like environment. Several outcomes are possible from this trial run. If performance improves in the expected direction, the experts may choose to stop tuning or test more aggressive configurations to seek further gains. If performance worsens, the experts will revisit and revise their previous decisions and try again. These failure cases are valuable learning opportunities, helping experts refine their understanding. Regardless of whether tuning is successful or not, human experts can summarize their experiences and distill them into “tuning knowledge” for that specific HPC platform. This accumulated knowledge becomes extremely valuable in reducing the number of tuning trials needed in the future.

While the approach that human experts take for tuning tasks differs greatly from existing autotuning methods, the outcomes are compelling. In practice, it is 
%unlikely that a human expert needs thousands of trail-and-errors to tune an application. It is 
common for them to arrive at a \textit{near-optimal configuration} within \textit{a handful of attempts}. 
%In fact, our I/O experts typically stopped tuning within five tries across all evaluation scenarios.
%We believe three key factors contribute to the exceptional efficiency of human experts in the autotuning process: \circled{1}) leveraging accurate knowledge about the system, \circled{2}) conducting test runs to gather real-world feedback, and \circled{3}) accumulating tuning knowledge via reflection and reasoning.
{Such efficient tuning requires synthesis of expert knowledge of system parameters, application I/O workloads, hardware specifications, and empirical evaluations -- a rare and manpower-intensive combination of skills. This is challenging for traditional autotuning frameworks to deliver but an ideal proving ground for agentic large language models (LLMs), which have the potential to autonomously employ diverse domain knowledge, conduct reasoning, and utilize external tools to address complex high-level tasks in a way that was not previously possible.}

In this study we present \texttt{STELLAR}, a \underline{S}torage \underline{T}uning \underline{E}ngine \underline{L}everaging \underline{L}LM's \underline{A}utonomous \underline{R}easoning, which reproduces exceptional tuning efficiency similar to that of human experts. \texttt{STELLAR} can be used by domain scientists to achieve \textit{near-optimal} I/O performance for their applications within a \textit{single-digit} number of attempts. 
%address the previous three key factors that contribute to the exceptional efficiency of human experts via the following key ideas.

%Recent rapid developments in Large Language Models (LLMs) have shown great promise in enabling this vision.
\texttt{STELLAR} consists of three main innovations to deliver this goal.
First, \texttt{STELLAR} leverages retrieval-augmented generation (RAG)~\cite{lewis2020retrieval} to accurately integrate into the tuning process domain knowledge such as file system manuals and cluster hardware specifications. Second, \texttt{STELLAR} uses external tools to interact with the real world. Specifically, \texttt{STELLAR} uses an LLM to autonomously analyze I/O trace logs (e.g., Darshan logs~\cite{darshan}) based on its needs, conduct test runs, and gather real-world feedback by interfacing with the HPC cluster. Third, \texttt{STELLAR} organically integrates reasoning capabilities, in-context learning~\cite{salewski2023incontext}, and chain-of-thought prompting~\cite{wei2023chainofthoughtpromptingelicitsreasoning} of LLMs to generate configurations, reflect on past attempts, summarize their experiences, and distill valuable insights. As \texttt{STELLAR} is applied to more applications, it continuously accumulates new tuning \textit{rule sets}, which can then be used to tune new applications with improved efficiency.
%Via carefully designing the workflows, we minimize the hallucinations and unstable reasoning. 
%This enables \texttt{STELLAR} to accumulate these distilled experiences into a consistent \textit{tuning rule sets}, which can then be used to tune new applications with continuously improved efficiency.

%Driven by this pressing need, STELLAR adopts a fundamentally different approach. Instead of using LLMs merely as a “Manual Reader” or “Forum Searcher” to narrow the search space, STELLAR relies entirely on enhanced Large Language Models to achieve the goal of tuning real applications in minimal number of tries. Specifically, STELLAR does not depend on online forums or published papers as tuning knowledge sources, since such information can be inaccurate, biased, or not tailored to the specific HPC cluster in use. Instead, STELLAR builds its foundation on file system manuals, which are significantly more accurate, and augments this with tuning knowledge accumulated through direct interaction with the target cluster. This allows STELLAR to leverage highly reliable and cluster-specific information for tuning. To implement such an idea, STELLAR includes the following three key components: (1) a RAG-based domain knowledge acquisition module, (2) a tool-based trace analysis and test-run executor, and (3) a reasoning-driven configuration generator and knowledge accumulator. 

Our evaluations show that \texttt{STELLAR} significantly improves the I/O performance of various benchmarks and real applications, achieving up to 7.8x speedup compared with the default. The entire tuning typically finishes within five attempts, even without any accumulated tuning knowledge. The resulting I/O performance also is comparable to, or even surpasses, what human experts can achieve.
More importantly, we demonstrate that the full version of STELLAR, augmented with tuning knowledge accumulated from prior tuning experiences on simple benchmarks, can consistently achieve near-optimal performance (compared with expert tuning) with less than five attempts when applied to new, previously unseen applications. These results highlight the promise of \texttt{STELLAR}, particularly its autonomous agentic LLM design, in tuning complex parallel file systems. The main contributions of this study are threefold:
\begin{itemize}[topsep=2pt, partopsep=0pt, itemsep=2pt, parsep=0pt]
    \item We propose the first LLM-based tuning engine for parallel file systems in HPC environments and show its effectiveness via extensive evaluations.
    \item We introduce a novel framework to combine general domain knowledge (from system manuals) with cluster-specific knowledge (from accumulated tuning rules) to efficiently tune I/O systems within a small number of iterations.
    \item We present an effective agentic LLM workflow that intelligently interacts with the system and distills tuning rules through iterative feedback.
\end{itemize}

%\dai{Specifically, GPTuner uses LLMs to filter the tunable parameters by parsing and understanding manuals and online forums, then using traditional optimization method (Bayesian Optimization) to converge the optimal configurations. DB-BERT basically follows the same idea while using Reinforcement Learning to find the near-optimal configurations.} 
%\texttt{STELLAR} differs significantly from them. It is the first LLM-based tuning engine specifically designed for parallel file systems to support parallel scientific applications. While GPTuner and DB-BERT have successfully reduced the number of tuning iterations from thousands to fewer than 100, the significantly higher cost of running parallel HPC applications, as required for tuning parallel file systems, compared to running micro benchmarks (\texttt{TPC-H} or \texttt{TPC-C}) for local database tuning, makes these existing methods still unacceptable in HPC environments.

The rest of the paper is organized as follows. In \S\ref{sec:background} we discuss relevant backgrounds of LLMs. \S\ref{sec:related} discusses the closely related work. In \S\ref{sec:design} we describe the architecture of \texttt{STELLAR} in detail. We present the extensive experimental results in \S\ref{sec:eval}. In \S\ref{sec:conclude} we present our conclusions and discuss future work.

\section{Background}
\label{sec:background}
In this section, we provide the necessary background for understanding HPC storage tuning and its main challenges. We also briefly introduce LLMs, LLM agents, and the key challenges of using LLMs to address complex system optimization problems.
%, to provide context for our design.
%In this section, we provide a background for storage system tuning to position our approach and the strengths and weaknesses of modern LLMs in the context of parameter tuning. Additionally, we compare LLM-driven tuning systems in other domains to differentiate our approach from related methods.

\subsection{HPC Parallel File System Tuning}
%Parallel file systems (PFSes) have become a critical component of HPC systems, serving data accesses from scientific applications. Delivering optimal performance is therefore essential. 
Modern parallel file systems often provide a large number of configurable parameters (or “levers”) that enable customizing behavior to meet the needs of different applications and HPC platforms. For instance, Lustre, one of the most widely used PFSs, exposes more than 150 tunable parameters~\cite{lustre_manual}, while Ceph, another popular PFS, includes thousands of parameters~\cite{lyu2020sapphire}. The sheer number of parameters, combined with their wide range of possible values, makes identifying the optimal configuration extremely challenging.
%Such a huge number of parameters places significant challenges in tuning the parallel file systems.

\subsubsection{Tunable Parameter Importance}
Not all parameters have the same impact on I/O performance. Some parameters are set before the parallel file system is mounted, such as \textit{mount\_point} and \textit{mount\_block\_size}, and are therefore not considered tunable at runtime. Other parameters control specific functionalities, and their values should be determined based on user needs rather than solely for improving I/O performance. For example, in Lustre, the parameters \textit{llite\_checksums} and \textit{osc\_checksums} enable or disable checksum mechanisms at the \textit{llite} and \textit{osc} layers, respectively~\cite{braam2019lustre}. While both significantly affect I/O performance, they should not be tuned just for performance gains; instead, their configuration should be guided by data integrity requirements. Additionally, some parameters are I/O related but have minimal impact on performance. For instance, Lustre’s \textit{lru\_size} parameter controls the number of client-side locks in the LRU cached locks queue. While this may optimize performance, it primarily affects memory usage rather than directly impacting I/O performance~\cite{george2021understanding}. 

What truly matters for tuning are the parameters that are both tunable at runtime and have a significant impact on I/O performance. For example, Lustre uses \textit{stripe\_size} and \textit{stripe\_count} to define file layout, which in turn dictates how I/O accesses will be distributed across available resources~\cite{george2021understanding}.
%determine how files are accessed in parallel across storage targets, which can significantly affect I/O performance~\cite{}. 
Parameters such as \textit{max\_\allowbreak rpc\_\allowbreak in\_\allowbreak flight} and \textit{max\_\allowbreak pages\_\allowbreak per\_\allowbreak rpc}  control the concurrency and size of data transfers, directly influencing both latency and bandwidth~\cite{qian2017configurable}.
%Therefore, for parallel file system tuning,
It is critical to identify this category of high-impact, tunable parameters and focus on them, rather than attempting to tune all parameters in a brute-force manner. \texttt{STELLAR} leverages a RAG mechanism to accurately extract tunable parameters. More details are discussed in \S\ref{sec:design}.

\subsubsection{I/O Patterns and Profiling}
Tuning a parallel file system should be performed on a per-application basis, since different applications often exhibit distinct I/O patterns and thus respond differently to the same parameter values.
Most parallel file systems support such tuning. For instance, Lustre's \textit{stripe\_size} and \textit{stripe\_count} can be set for individual files. Its \textit{llite.*}, \textit{osc.*}, \textit{mdc.*}, and many other client-side parameters can be configured differently across compute nodes, affecting individual applications.

Understanding the internal I/O behavior of applications is hence critical for tuning. In HPC environments, multiple I/O profiling tools have been developed, such as STAT~\cite{STAT}, mpiP~\cite{mpip}, IOPin~\cite{IOPIN}, Recorder~\cite{Recorder}, and Darshan~\cite{darshan}. Each of these tools provides insights into various aspects of application I/O behavior. In this study, we leveraged Darshan logs for two primary reasons. First, Darshan is lightweight and requires no application modification, making it easy to apply to a broad range of applications~\cite{moti2023trace}. For the same reason, it is widely installed across many HPC facilities~\cite{nersciotuning2025}. Second, several recent studies using LLMs to analyze I/O behavior~\cite{egersdoerfer2024ion,drishti} have shown that Darshan logs can be efficiently chunked and distilled into high-level, actionable summaries. Specifically, Darshan traces key statistical metrics for each file across different types of I/O interfaces, including POSIX I/O, MPI-IO, and Standard I/O. These metrics include the amount of read/write data, aggregate time for read/write/meta operations, the ID of the rank issuing I/O requests, and the variance of I/O size and time among different application ranks. More details about how we process Darshan logs will be discussed in later sections.
%Darshan also collects Lustre file system-level metrics such as stripe width and OST IDs over which a file is striped. 

%traces unique sets of counters to characterize the behavior of various modules across the HPC I/O stack, including POSIX (Portable Operating System Interface) I/O, MPI (Message Passing Interface) I/O, Standard I/O, as well as many other extensions. The set of counters provided by Darshan for each module captures characteristics primarily related to the volume of data and volume of operations passed through the module organized by MPI rank and file targets. 
%In the context of Storage System tuning, leveraging such information is vital to enabling rapid convergence towards optimal parameter settings. This is due to the fact that the optimal range of settings to explore for a given set of tunable storage system parameters is primarily determined by the way in which the application conducts I/O with the storage system (its I/O behavior). 

%%% Plot Position
\begin{figure*}[h!]
    \centering
\includegraphics[width=0.8\textwidth]{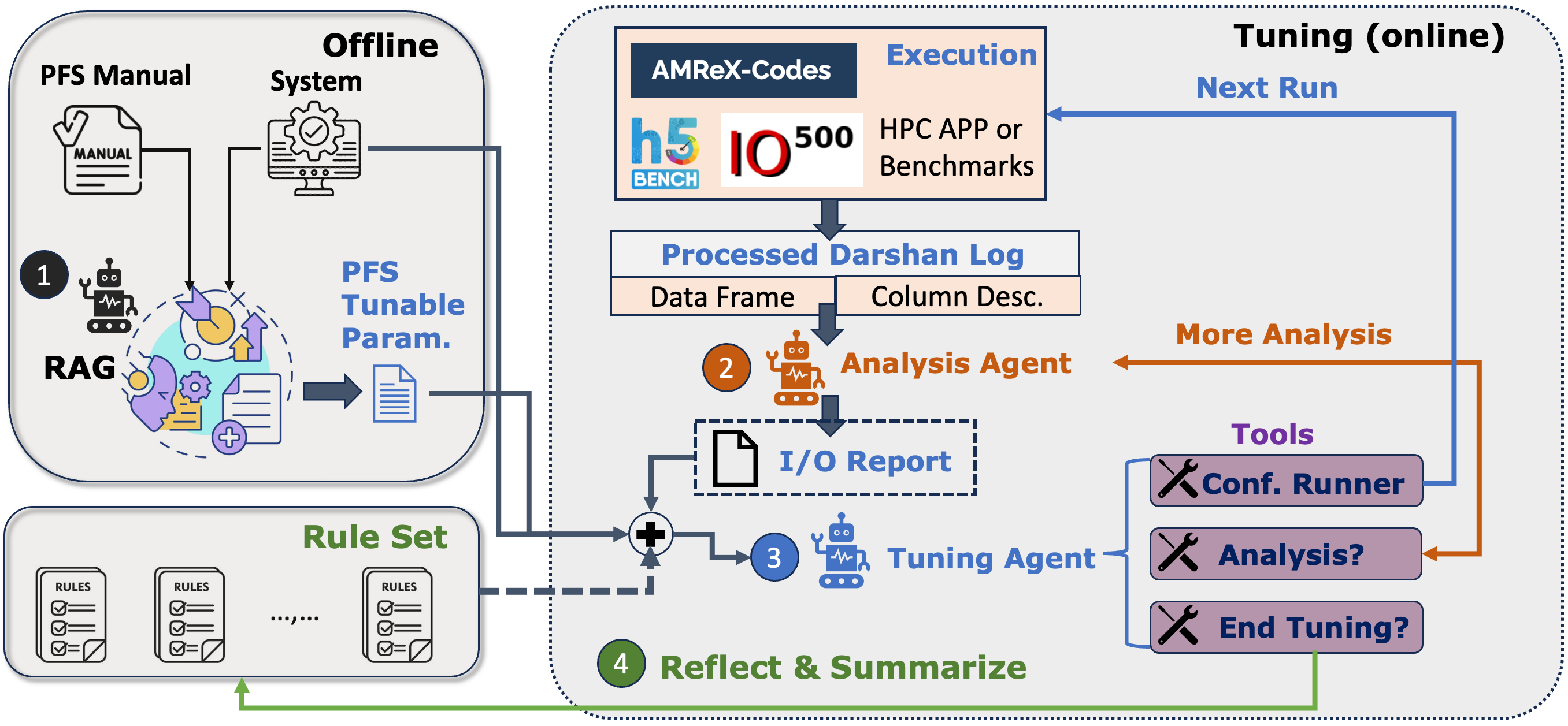}
    \caption{STELLAR design overview. The four numbered elements represent the four key modules in \texttt{STELLAR}.}
    \label{STELLAR-design}
    \vspace{-1em}
\end{figure*}
%%% Plot Position
\subsection{Large Language Models}
The rapid development of LLMs has demonstrated their powerful capabilities across a wide range of tasks. The latest models, such as GPT-4.5~\cite{openaiGPT45} and Claude 3.7~\cite{anthropicClaude37}, further showcase their strengths in handling complex and in-depth tasks, such as programming or conducting scientific surveys~\cite{openaideepresearch2025}. 

Recently, LLMs have entered a new era with the emergence of LLM agents, software components powered by LLMs that are capable of perceiving environments, reasoning about goals, and executing actions~\cite{luo2025large}. Unlike traditional chatbots, LLM agents can interact with real systems through continuous exploration, reasoning, and adaptation, as demonstrated by systems such as DeepResearch~\cite{openaideepresearch2025}, OpenAI Operator~\cite{openaiOperator}, and Manus~\cite{monicaManus}. Various tools are available for developing LLM agent systems, including AutoGen~\cite{autogen2025}, LangChain~\cite{langchain2025}, and Dify~\cite{dify2025}. Numerous LLM agents have been developed across scientific domains, such as SciAgents~\cite{ghafarollahi2024sciagents}, Curie~\cite{kon2025curie}, ChemCrow~\cite{bran2023chemcrow}, and AgentHospital~\cite{li2024agent} among others~\cite{luo2025large}.

\subsubsection{Workflows and Agents.}
LLM-based agentic systems can be categorized into two types based on their level of autonomy: \textit{Workflows} and \textit{Agents} ~\cite{anthropicBuildingAgents2024}. \textit{Workflows} are systems where LLMs and tools are orchestrated through predefined code paths, whereas \textit{Agents} are systems in which LLMs dynamically direct their own processes and tool usage, maintaining control over how they accomplish tasks. Generally, workflows prioritize reliability at the cost of flexibility. They are well-suited for well-defined tasks where steps can be predetermined and the total number of execution paths is limited. In contrast, agents may compromise some reliability but offer the flexibility to address problems that developers did not anticipate from the outset. They are best applied to problems where future execution steps depend heavily on the outcomes of previous ones and where the enumeration of possible execution paths is large. It is therefore critical to make informed design choices when developing an LLM-based agentic system.

%Since the release of OpenAI's GPT-3.5 (ChatGPT), there has been a significant rise in multi-step autonomous software systems to arrive at some predetermined goal. This is primarily due to the robust instruction following capabilities adapted by LLMs through alignment training techniques applied after their general pre-training. However, a key distinction between many of these systems is whether they are designed to follow predetermined paths (workflows) or to autonomously control their own execution paths (agents) as the means to their goals. 
%Generally, the former prioritizes reliability at the cost of flexibility and the latter prioritizes flexibility at the cost of reliability, while the true cost of this trade off depends heavily on implementation details. As such, there are many scenarios in which LLM workflow approaches are the clear choice as well as the opposite. 

\subsubsection{Tool Usage}
Tool use is another critical component of LLM-based agents, enabling them to interact with their environment, for example by performing calculations, accessing real-time information, and generating code. Tool usage involves two key steps: deciding when to use a tool and choosing the right tool. The tool-use decision refers to determining whether a tool is needed to solve a given problem. When the agent is generating content with low confidence or encountering problems that require specific tool functionalities, it should decide to invoke the appropriate tool. Tool selection, on the other hand, involves understanding the available tools and assessing the agent’s current context to choose the most suitable one. A successful LLM-based agentic system must have a robust infrastructure to support both of these steps.

\subsubsection{Self-Learning}
Self-learning is key to continuously improving LLM agents’ capabilities. The typical methods are \textit{self-reflection} and \textit{self-correction}, both of which enable LLMs to iteratively refine their outputs by identifying and addressing errors. These methods leverage the self-reflective capabilities of modern LLMs, which emerged during training as the models were encouraged to check and reflect on their answers. It has been shown that promoting self-reflection instructions can significantly enhance an LLM agent’s performance~\cite{shinn2023reflexion}. When building a continuously learning LLM agent, it is therefore critical to effectively harness this property.

\subsubsection{Hallucination Issues of LLM Agents}
The knowledge of LLM agents remains limited in highly domain-specific areas, such as tuning the parameters of a particular parallel file system. Although their training data may include manuals for these systems and relevant discussions from online forums, the rarity of such data often means that LLMs are not sufficiently trained to capture accurate information. In such cases, when asked relevant questions, LLMs are prone to hallucinate knowledge. Worse yet, without a known ground truth, these hallucinations are difficult to detect, as LLMs often present them with authoritative language, potentially affecting downstream tasks in LLM workflows or agent systems. It is therefore critical to address the hallucination issue when implementing a practical tuner.

%Although hallucination is still an inherent issue of LLMs, via  have become very reliable sources of knowledge when operating in commonly accessible and well-documented domains. This is due to the fact that LLMs are trained by being rewarded for accurate next-token prediction. In such a setup, it is easily understood that information about domains more commonly included in their training data carry a more significant portion of the total possible reward and therefore are more likely to be captured in their weights. However, in domains where the amount of training data is limited, it is well known that LLMs often hallucinate knowledge. Worse yet, without a known ground truth, hallucinations are difficult to detect as LLMs use similar authoritative expression when generating either fact or fiction. 

\section{Related Work}
\label{sec:related}

We consider two threads of work that are most relevant with \texttt{STELLAR}: (1) the autotuning frameworks built for high-performance parallel file systems and (2) the autotuning frameworks that leveraged LLMs but are designed for other storage systems.

\subsection{Autotuning for HPC Parallel File Systems}
%Auto-tuning systems have been widely studied and proven complicated to explore the large optimization space~\cite{dorier2022hpc,bernstein2002complexity}. 
%The autotuning approach for system optimization can be framed as a mixed-integer nonlinear optimization problem with a computationally intensive black-box objective function~\cite{dorier2022hpc}. Introducing a dynamic component to autotuning transforms it into a Decentralized Partially Observable Markov Decision Process, a problem type that has been classified as NEXP-hard~\cite{bernstein2002complexity}. Ideally, to generate an optimal solution, a tuning system would need access to a complete history of observations, encompassing both past and future requests; however, such comprehensive data is practically unachievable. Consequently, all existing tuning strategies have resorted to implementing some form of approximation in their methodologies~\cite{li2017capes}, acknowledging the inherent limitations while striving for optimal system performance.
Autotuning HPC parallel file systems began from using heuristic-based approaches. For instance, ASCAR~\cite{li2015ascar}, TAPP-IO~\cite{neuwirth2017automatic}, DCA-IO~\cite{kim2019dca}, and IOPathTune~\cite{rashid2023iopathtune} introduced rule-based strategies to tune parameters such as stripe size and stripe count to handle burst and imbalance I/Os. They share the same limitations in handling dynamically varying workloads, which motivate the machine learning-based autotuning that learns tuning strategies directly from the data samples. SAPPHIRE~\cite{lyu2020sapphire} utilizes Bayesian optimization to recommend the best configurations. Behzad et al.~\cite{behzad2019optimizing} implemented nonlinear regression models to model I/Os and employed genetic algorithms to explore the configuration space, which were similarly used by Rajesh et al.~\cite{tunio}. Cao et al.~\cite{cao2018towards} further conducted a comprehensive comparative study across multiple autotuning strategies to assess their efficiency. 
CAPES~\cite{li2017capes}, AIOC2~\cite{cheng2021aioc2}, and Magpie~\cite{zhu2022magpie} leveraged deep reinforcement learning learning to adaptively learn tuning strategies. 
However, the high costs of iteratively sampling and testing workloads to explore the search space make these methods prohibitively expensive, limiting their usage at scale. %\texttt{STELLAR} is the first LLM-based autotuner designed to address such an issue for HPC systems.

\subsection{LLM-Based Database Tuning}
Local storage systems, such as databases, face tuning complexities similar to those parallel file systems face. A variety of autotuning approaches, such as heuristic-based, machine learning-based, reinforcement learning-based, and even LLM-based, have been explored in the database context~\cite{duan2009tuning,van2017automatic,zhang2019end}. However, because of substantial differences between these two domains in terms of workloads (e.g., SQL queries vs. scientific applications), storage engines (e.g., SQL or key-value stores vs. parallel file systems), and hardware specifications (e.g., storage nodes vs. HPC clusters), it is not feasible to apply frameworks developed for one domain to the other without significant redesign. This distinction is also evident in the literature, where autotuning methods for SQL databases are typically not compared directly with those for parallel file systems. Following this established practice, we also do not quantitatively compare \texttt{STELLAR} with database-focused methods in our evaluations.

It is still insightful, however, to conceptually compare \texttt{STELLAR} with recent methods that also leverage LLMs for tuning database systems. Notable examples include DB-BERT~\cite{trummer2022db}, GPTuner~\cite{lao2023gptuner}, and E2ETune~\cite{huang2025e2etuneendtoendknobtuning}, which represent some of the most recent and representative efforts in this space. DB-BERT fine-tunes pretrained language models (i.e., BERT models) to translate natural language hints into configuration recommendations, followed by reinforcement learning to iteratively refine the initial selections. Similarly, GPTuner employs LLMs to extract tuning parameters and candidate values from manuals and online forums, then applies Bayesian optimization to converge on optimal configurations. E2ETune avoids the costly online iterative phase by shifting it to an offline data collection stage, where workloads are still repeatedly executed to gather samples using a Gaussian process. Notably, all these approaches successfully reduce the number of required iterations from thousands to fewer than 100. In the HPC context, however, the significantly higher cost of running parallel scientific applications makes even this reduced iteration count prohibitively expensive. In contrast, \texttt{STELLAR} adopts a fundamentally different approach by fully leveraging agentic LLMs to perform informed optimization, enabling it to converge within a single-digit number of attempts.

\section{Design and Implementation} 
\label{sec:design}
\subsection{Overall Workflow}

Figure~\ref{STELLAR-design} shows the overall design of \texttt{STELLAR}. The design consists of two main parts that together form the complete \texttt{STELLAR} engine.
The \textit{Offline} part (shown on the left) runs before any tuning occurs. In this part, \texttt{STELLAR} implements a RAG-based process to extract parameters and their definitions. The RAG system takes the parallel file system manual as input. It then outputs a filtered list of high-impact tunable parameters, along with accurate descriptions for each parameter.
The \textit{Tuning (Online)} part (shown on the right) conducts the actual iterative tuning process. \texttt{STELLAR} begins tuning after an initial run of the target application (e.g., E3SM, H5Bench). This initial run generates a Darshan log, which is further processed into a set of Pandas \textit{DataFrames}, accompanied by a separate file describing the meaning of each column. The dataframes are sent to our first agent, the \textit{Analysis Agent}, to extract application-specific runtime I/O behaviors. The \textit{Analysis Agent} can autonomously write and execute analysis code to interpret processed Darshan logs based on specific requirements.

The I/O Report generated by the \textit{Analysis Agent}, together with the PFS Tunable Parameters retrieved during the offline phase, is sent to the \textit{Tuning Agent}, which drives the main trial-and-error loop. We have implemented three tools that the \textit{Tuning Agent} calls to perform specific actions. The \textit{Analysis Tool} is used to determine whether additional analysis is necessary. If so, it instructs the \textit{Analysis Agent} to generate new code for the required analysis. The \textit{Configuration Runner Tool} generates a new set of parameter values and drives the target application to run again with the updated parameter values. The \textit{End Tuning Tool} determines whether the trial-and-error loop should stop, thereby concluding the tuning process.

Once tuning concludes, we use the \textit{Reflect \& Summarize} module to reflect on the entire trial-and-error loop and summarize rules derived from the experience. These new rules are merged with any existing rules to form a comprehensive global \textit{Rule Set}. This global \textit{Rule Set} will be utilized for tuning future applications as part of the input context provided to the \textit{Tuning Agent}, as shown in Figure~\ref{STELLAR-design}. A complete \texttt{STELLAR} run, leveraging a non-empty global \textit{Rule Set}, ensures maximal tuning efficiency.

In the following subsections we introduce each individual component in greater detail.

%To enable detailed and application-specific extraction of runtime I/O behavior, we implement the \textit{Runtime Analysis Agent} as an agent, capable of autonomously writing and executing analysis code. To carry out the tuning process and allow STELLAR to test assumptions based on measurable performance changes, we implement the \textit{Tuning Agent} as an autonomous agent with access to three tools. Finally, once the \textit{Tuning Agent} ends the tuning process, it is asked to summarize a set of rules which were learned throughout the procedure, and these rules are merged with any existing set of rules in the \textit{Global Rule Synthesis} component. 

%We are workflow and Agent hybrid system. Describe them and list their pros/cons. Then, discuss why we pick the current design.

%For certain aspects, we do prefer fixed workflow, such as xxx. Explain why we think fixing this helps.

%For certain aspects, 1) darshan information extraction, 2) when to stop the tuning. Discuss why these two cases will be better with agent design. 

\subsection{RAG-Based Parameter Extraction}
Successful and efficient parameter tuning for a parallel file system relies heavily on two key pieces of information: (1) a curated list of parameters significantly impacting I/O performance and (2) accurate descriptions of these tunable parameters, including how they affect I/O performance and their valid ranges. This detailed information helps the \textit{Tuning Agent} generate correct parameter settings. In this subsection we discuss how \texttt{STELLAR} extracts these parameters.

\subsubsection{Limitations of LLMs in Extraction}
Although LLMs have demonstrated remarkable abilities to answer various difficult questions, they still suffer from hallucinations — plausible but factually incorrect outputs. This issue is particularly problematic in our context because parallel file systems are domain-specific and lack extensive amounts of organized online documentation and highly popular discussion boards.

\begin{figure}[t]
    \centering
\includegraphics[width=0.9\columnwidth]{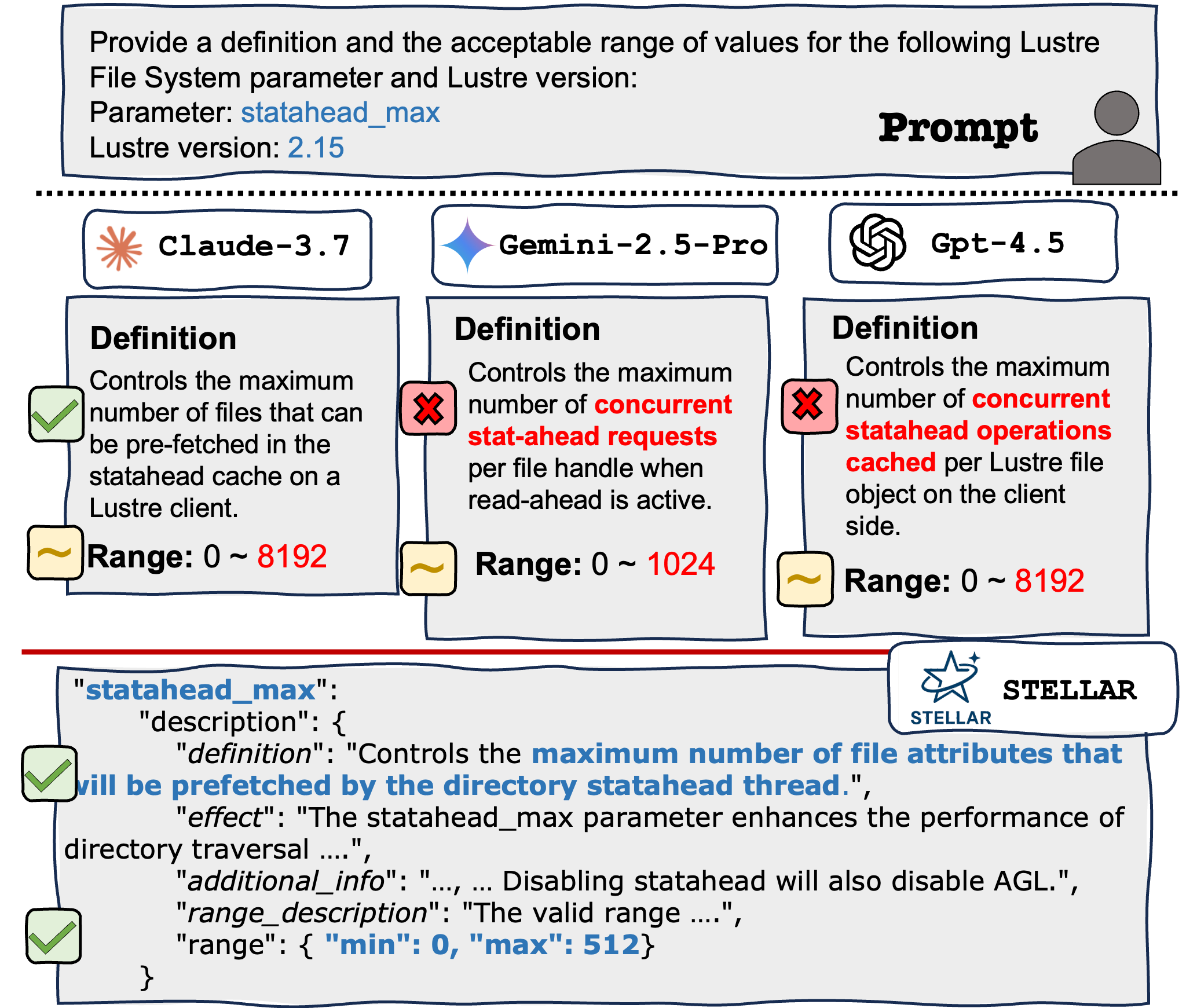}
    \caption{Example of LLM hallucinations for storage system parameter details. We also show the RAG-based extraction result of \texttt{STELLAR} on the same parameter. Note that our RAG-based extraction leverages the older GPT-4o model.}
    \label{hallucinations}
    \vspace{-1em}
\end{figure}

We illustrate such hallucinations in practice in Figure \ref{hallucinations}, where three state-of-the-art LLM models provide definitions and accepted ranges for the \textit{statahead\_max} parameter in Lustre version 2.15. Here, a check mark indicates a correct result, a cross an incorrect result, and a tilde an imprecise result. None of the three models provided entirely correct responses. All three were incorrect regarding the maximum accepted value for the parameter, and both OpenAI's GPT-4.5 and Google's Gemini-2.5-Pro provided flawed parameter definitions. 
We further show the output of \texttt{STELLAR} on the same parameter at the bottom of the figure. We can see that not only does it provide more detailed aspects about this parameter but it also generates correct information, which results from the RAG-based parameter extraction design.

Directly feeding entire manuals into language models expecting accurate parameter descriptions is also problematic due to the limitations of current LLMs. Manuals, such as the Lustre manual \cite{lustre_manual} with over 600 pages (>300k tokens), often exceed typical LLM context windows. 
%dong: good edits. I accepted it. \textcolor{red}{Even models supporting extensive contexts encounter problems such as \textit{lost-in-the-middle truncation}\cite{lostinmiddle}, which neglects central context information, and \textit{losing long-range dependencies}\cite{borgeaud2022improvinglanguagemodelsretrieving}, limiting effective long-document modeling.}
\textcolor{black}{Even models supporting extensive contexts encounter problems such as \textit{lost-in-the-middle truncation}\cite{lostinmiddle}, \textit{losing long-range dependencies}\cite{borgeaud2022improvinglanguagemodelsretrieving}, and \textit{context rot}\cite{hong2025context}, limiting their ability to extract and combine key pieces of relevant information scattered throughout large documents.}

\subsubsection{RAG-Based Extraction}
\texttt{STELLAR} addresses the hallucination issue using a RAG-based workflow. Instead of directly feeding manuals into LLMs, we first build a \textit{vector index} from the system manual, then implement a multistep filtering mechanism that leverages retrieved information to identify key parameters and generate accurate descriptions.

\textit{\textbf{Generating the vector index.}} We create an embedding index by chunking and embedding the entire parallel file system manual using LlamaIndex \cite{Liu_LlamaIndex_2022}, a popular open-source embedding and retrieval framework. We use the default chunk size of 1,024 tokens, 20-token overlap, and OpenAI's \textit{text-embedding-3-large} model \cite{text-embedding-3-large} to embed each chunk.
%
%\textcolor{red}{The \textit{vector index} effectively extracts relevant information chunks about tunable parameters scattered throughout the manual into a queryable database.} 
\textcolor{black}{The \textit{vector index} stores the text chunks extracted from the parallel file system manual in a queryable database}. Queries to the \textit{vector index} then retrieve concise and highly relevant context for the LLMs to further analyze, 
%\textcolor{red}{directly addressing both truncation and long-range dependency issues}
\textcolor{black}{thereby mitigating the risks associated with long contexts}. Additionally, this index is easily updated when new manual versions become available.

\textit{\textbf{RAG-based parameter definition.}}
To facilitate runtime configuration, modern parallel file systems typically expose standard interfaces to access tunable parameters. These can serve as an initial source for parameter identification. For example, Lustre exposes parameters under \textit{/proc/fs/} and \textit{/sys/fs/}. Initially, a rough filter selects only writable parameters since these can be altered by \texttt{STELLAR}. We note that this step may not always be necessary because some storage systems directly expose tunable parameters via configuration files (e.g., DAOS~\cite{hennecke2020daos}).
%For Lustre such initial filtering rules out [add evidence from lustre]
%a large number of parameters. However, in some cases where parameters are mixed with statistics and other indicators (i.e.

For each parameter, we query the \textit{vector index} with the question \textit{"How do I use the parameter [parameter name]?"}, retrieving the top \textit{K} (e.g., 20) relevant chunks, leveraging the default LlamaIndex retriever. Using the retrieved chunks, an LLM (defaulting to OpenAI's GPT-4o \cite{openai2024gpt4o}) is prompted to determine whether the documentation provides sufficient information to define the parameter's purpose and valid range. If the documentation is sufficient, the LLM is prompted to describe the parameter's purpose, its intended impact on I/O, and specify its valid range. An example output was shown in Figure~\ref{hallucinations}.
% based on cosine similarity of the text chunks and the embedded representation of the question
%Given the results of the query, we prompt LLMs with the retrieved text chunks and ask 1) indicate if the documentation provides sufficient information about the given parameter, 2) provide how the parameter controls I/Os along with any documented effects of tuning the parameter, and 3) the acceptable value range as documented in the documents. %any additional information useful to describe when the parameter should be used, 
%By default, \texttt{STELLAR} uses OpenAI's GPT-4o \cite{openai2024gpt4o} for such a task. 
Parameters that are found to have insufficient documentation are filtered out based on the assumption that parameters that are not described in the documentation are likely to be of lesser importance than those that are. 

Notably, in many cases a parameter may depend on other parameters, complicating this procedure. For instance, in Lustre the maximal value of \textit{max\_\allowbreak read\_\allowbreak ahead\_\allowbreak per\_\allowbreak file\_\allowbreak mb} is half of \textit{max\_\allowbreak read\_\allowbreak ahead\_\allowbreak mb}, whose maximal value is half of the system memory. For those parameters, deciding the range involves calculation and hardware specifics. In order to handle them, the LLM is instructed to use specific \textit{dependent} and \textit{expression} syntax rules that can be parsed and evaluated in the online tuning setting. These expressions will be calculated based on actual system values during tuning.
%Note that, expressions containing only numeric values and system variables are always evaluated while those containing dependencies to other parameters are only evaluated when depended parameters are not in the final filtered set of parameters being tuned.

\textit{\textbf{Selecting important parameters.}}
After obtaining a filtered set of accurately described parameters with defined ranges, we select the most impactful ones since not all parameters equally influence I/O performance.
%We need to analyze the remaining parameters to extract a subset which is most likely to have an impact on performance. 
To do so, we first exclude the binary parameters because their settings typically represent user trade-offs rather than tuning decisions. For example, the binary checksum parameters in Lustre significantly impact I/O performance~\cite{fragalla2014configure,tanaka2019performance} but risk data integrity, a tradeoff best left to users.

%review all of the information extracted via the previous RAG-based definition step to 
We then prompt the LLM (e.g., GPT-4o) to decide, with documented reasoning, whether each remaining parameter is likely to have a significant impact on performance. 
This is feasible because the parallel file system manual typically describes how each parameter changes the I/O behavior from which potential to impact performance can be inferred.
%human operators base their decisions regarding which parameters are worth tuning on their reasoning over what the parameter is intended to control and intended effect. 
%todo: \textcolor{red} {do we have an image of how this works? it would be helpful. But i think we do not have enough space.}
For example, Lustre's \textit{max\_rpcs\_in\_flight} parameter controls the maximum number of concurrent remote procedure calls (RPCs) between object storage clients (OSCs) and object storage targets (OSTs), clearly impacting I/O performance.
In contrast, Lustre parameters such as \textit{nrs\_\allowbreak delay\_\allowbreak min}, \textit{nrs\_\allowbreak delay\_\allowbreak max}, and \textit{nrs\_\allowbreak delay\_\allowbreak pct} simulate high server load scenarios, which is relevant but not directly connected to I/O performance. LLMs such as GPT-4o can make reliable selections by leveraging these detailed parameter descriptions.
%Leveraging detailed parameter descriptions from documentation like these, LLMs such as GPT-4o can make reliable selections. 
%LLMs can similarly decide that they should not be chosen as high impact parameters. Given the detailed information extracted from the documentation for each parameter, such decisions can be reliably made by LLMs .

The final output, \textit{PFS Tunable Parameters}, is provided to the \textit{Tuning Agent} as discussed next.  The final set of parameters is likely to be much smaller than the complete set of parameters. For Lustre, \texttt{STELLAR} chooses a subset of 13 parameters to tune.
%from the original available in the Lustre file system.
%For the Lustre file system, the final set of parameters is significantly smaller than the initial set of parameters found using lctl, including only 13 total parameters.

\subsection{Agentic Online Tuning}
The online tuning process in \texttt{STELLAR} is implemented as a fully autonomous agentic system involving the \textit{Analysis Agent} and the \textit{Tuning Agent}. There are essentially two loops during online tuning. The first is the main trial-and-error loop, which includes executing the target application, analyzing Darshan logs, generating a new set of configurations, and looping back to execute the application again with a different set of parameter values. The second is the minor loop, where the \textit{Analysis Agent} generates an I/O Report, and the \textit{Tuning Agent}, upon finding the report incomplete, requests additional analysis from the \textit{Analysis Agent}.

\subsubsection{Analysis Agent}
The \textit{Analysis Agent} serves two key purposes. First, it provides the context for the iterative tuning process by analyzing I/O behavior from Darshan logs collected at runtime and summarizing its findings. Its secondary role is conducting any additional specific analyses requested by the \textit{Tuning Agent}. Both roles are enabled by designing the \textit{Analysis Agent} as a code-executing LLM agent, leveraging the OpenInterpreter~\cite{openinterpreter} framework for autonomous task completion. %\textcolor{blue}{. Specifically, the \textit{Analysis Agent} leverages the OpenInterpreter \cite{openinterpreter} framework, which enables the agent to execute generated code in a Jupyter kernel environment}. 
Given a task description, the agent plans, implements, and executes code until the task is considered complete by the agent.

The \textit{Analysis Agent} operates on Pandas \textit{DataFrames} along with string variables describing columns, rather than raw logs. Our preprocessing script extracts counters for each module (e.g., POSIX, MPI-IO) from Darshan and loads them into separate dataframes with corresponding counter descriptions. The log header is also loaded as a string variable. This process can be replicated for other tracing frameworks such as Recorder or less granular sources such as Elbencho \cite{elbencho}, which reveal useful details about I/O behavior.

The \textit{Analysis Agent} is tasked with providing a high-level summary of the application's I/O behavior by inspecting loaded variables (dataframes and descriptive strings), identifying files accessed by the application, and highlighting any information it deems useful for tuning the parameters. 
This high-level task description allows for dynamic analysis where the agent decides the most appropriate analysis based on application context. 
Although this approach may overlook certain details, the \textit{Tuning Agent} can request further specific analysis during the iterative tuning process when necessary.
%\textcolor{blue}{add prompt outline or figure if space allows}
%Doing so may miss some rather critical details. We allow the \textit{Tuning Agent} to ask for further specific analysis when it deems it to be necessary.

\begin{figure}[t]
    \centering
\includegraphics[width=0.8\columnwidth]{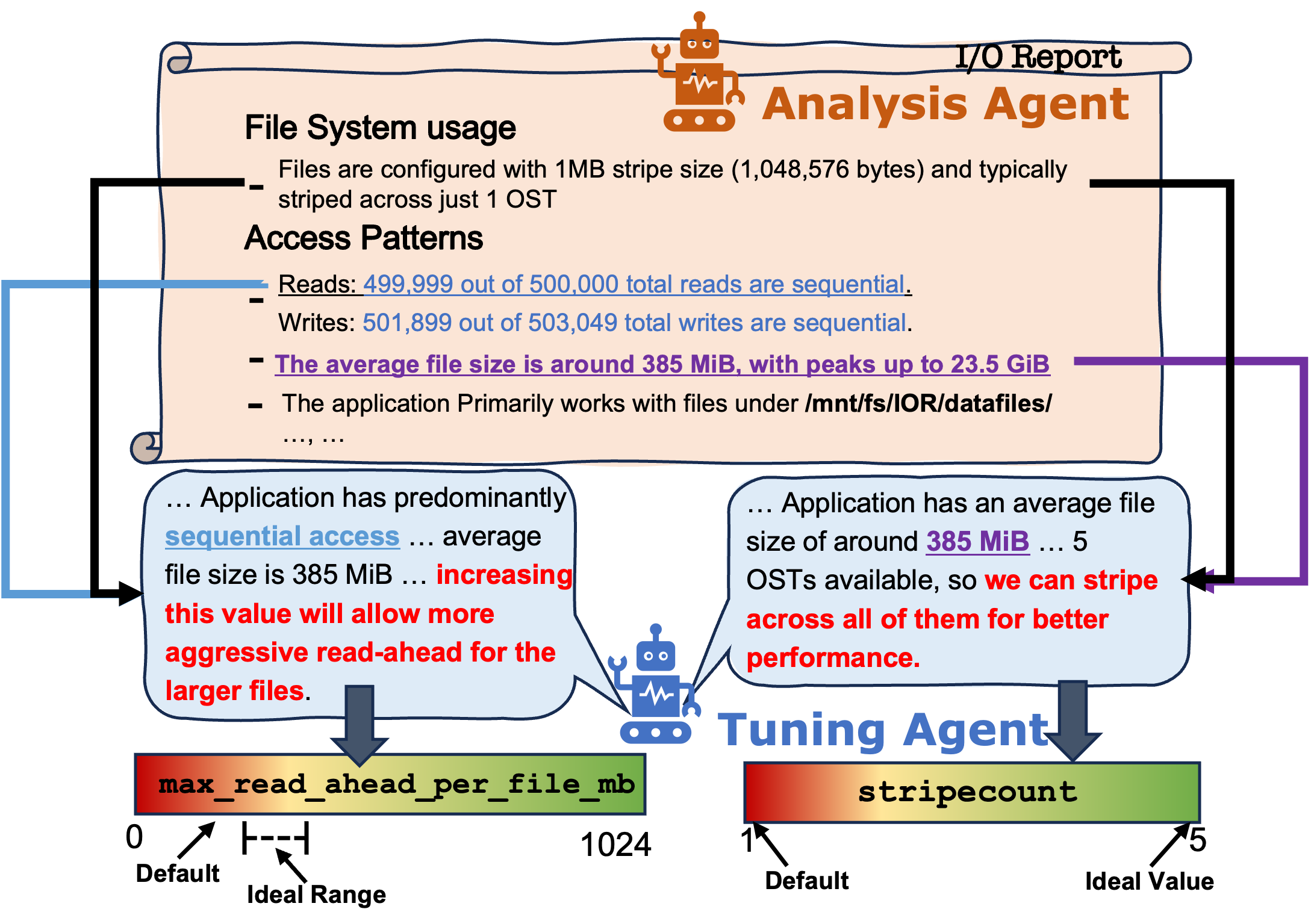}
    \caption{Example of decision-making via interactions between the \textit{Analysis Agent} and \textit{Tuning Agent}.}
        \vspace{-1em}
    \label{decision-sample}
\end{figure}

\subsubsection{Tuning Agent}
%the I/O report of the target application, details of the hardware and storage system setup, a
The \textit{Tuning Agent} is the primary controller of the iterative tuning process. Its goal is to generate high-quality configurations, observe actual application performance, and reflect on the outcomes. At the beginning of the tuning process, the \textit{Tuning Agent} receives the final filtered set of tunable parameters, details about the hardware and storage system setup, and the I/O Report generated by the \textit{Analysis Agent}.
%\textcolor{blue}{add prompt outline or figure if space allows} 
The \textit{Tuning Agent} then makes decisions using one of three potential environment interactions implemented as LLM tool calls. 

If the \textit{Tuning Agent} finds relevant information to be missing for the parameter it tries to tune, it selects the \textit{Analysis?} tool and formulates a specific question (prompt) for the \textit{Analysis Agent}.

If confident, the \textit{Tuning Agent} generates new configurations and executes the application to verify performance improvements. When generating new configurations, it explicitly documents the rationale behind each parameter value selection. This process encourages careful thought and facilitates validating LLM knowledge about parameter impacts by comparing stated reasoning against actual performance outcomes, which serves as the key to formulate \textit{Tuning Rules}. 
To run the target application, \texttt{STELLAR} requires details on the initial execution process and interactions with the scheduling system. These should be provided by the domain scientists.

%If the \textit{Tuning Agent} is confident. It will choose to generate the configurations and run the target application again to verify the performance improvements. For generating the new configurations, it needs to detail why certain value is selected for each parameter. We prompt the LLMs explicitly to generate the reasons, aiming to think more carefully and will use that reasons to cross-check with the actual performance change for validating if the LLMs have correct knowledge about certain parameters or not. For running the target application again, it needs the help to know how the initial execution was conducted. Such information needs to be provided to \texttt{STELLAR}. Additionally, how to interact with the scheduling system should also be provided to \texttt{STELLAR} should also be provided. 

When the \textit{Tuning Agent} selects the \textit{End Tuning?} tool, it must provide reasoning for this decision. Specifically, in the system prompt we instruct the agent to finalize the process only when it believes that further tuning would not elicit further performance gains and to provide justifications.
%describe in detail what was done during the process along with its reason for why that justifies ending the process. 
In practice, these guidelines cause the agent to explore more when significant performance improvement has not been found since it is less confident that improvement cannot be found elsewhere. %with more search and continue tuning until 
The agent typically stops when diminishing returns are reached given the performance has noticeably improved beyond the default performance. 
The \textit{End Tuning?} tool terminates the loop and initiates the \textit{Reflect and Summarize} step. Here, the agent synthesizes rules learned during tuning, describing optimal parameter settings based on observed application I/O behavior. This step enables offline knowledge-base generation, as discussed further in the subsequent section.

\subsubsection{Workflow vs. Agent?}
%based system where the set of tunable parameters, the initial measured application performance using the default parameter settings, and a high level I/O behavior analysis of the application are given as input to the \textit{tuning agent}. 
%The \textit{tuning agent} then acts as an autonomous agent with a limited set of permitted environment interactions including the ability to request specific I/O analysis, the ability to generate a new prediction for what the best parameter values should be, and the ability to conclude the tuning process. For the initial application I/O behavior summary and the option to ask for additional analysis during the tuning process, the \textit{Tuning Agent} relies on the separate \textit{Runtime Analysis Agent} described in more detail in the following section. 
As previously discussed, the online tuning in \texttt{STELLAR} uses a system of two fully autonomous agents instead of more controllable \textit{Workflows}, primarily because of the adaptive nature required at this stage. 
First, some applications with more complex behavior may be inherently less trivial to tune compared with others, requiring dynamic extension or shortening of the tuning process. 
Additionally, various applications have different I/O behavior that inherently requires a dynamic approach to summarizing their unique behaviors, one that is capable of dynamically adapting the focus of the summary to the most important aspects.

\subsection{Rule Set Accumulation}
Following the conclusion of every tuning procedure controlled by the \textit{Tuning Agent}, the agent is asked to summarize, in the form of a rule set, what has been learned during the process. The first time \texttt{STELLAR} runs on a given system, the global \textit{Rule Set} is empty. In subsequent runs, a new rule set will be added to the initial context and updated with new information learned from each run. Following this process, it continuously accumulates more rules as \texttt{STELLAR} is used to tune more applications.

\begin{figure}[ht!]
    \centering
\includegraphics[width=0.8\columnwidth]{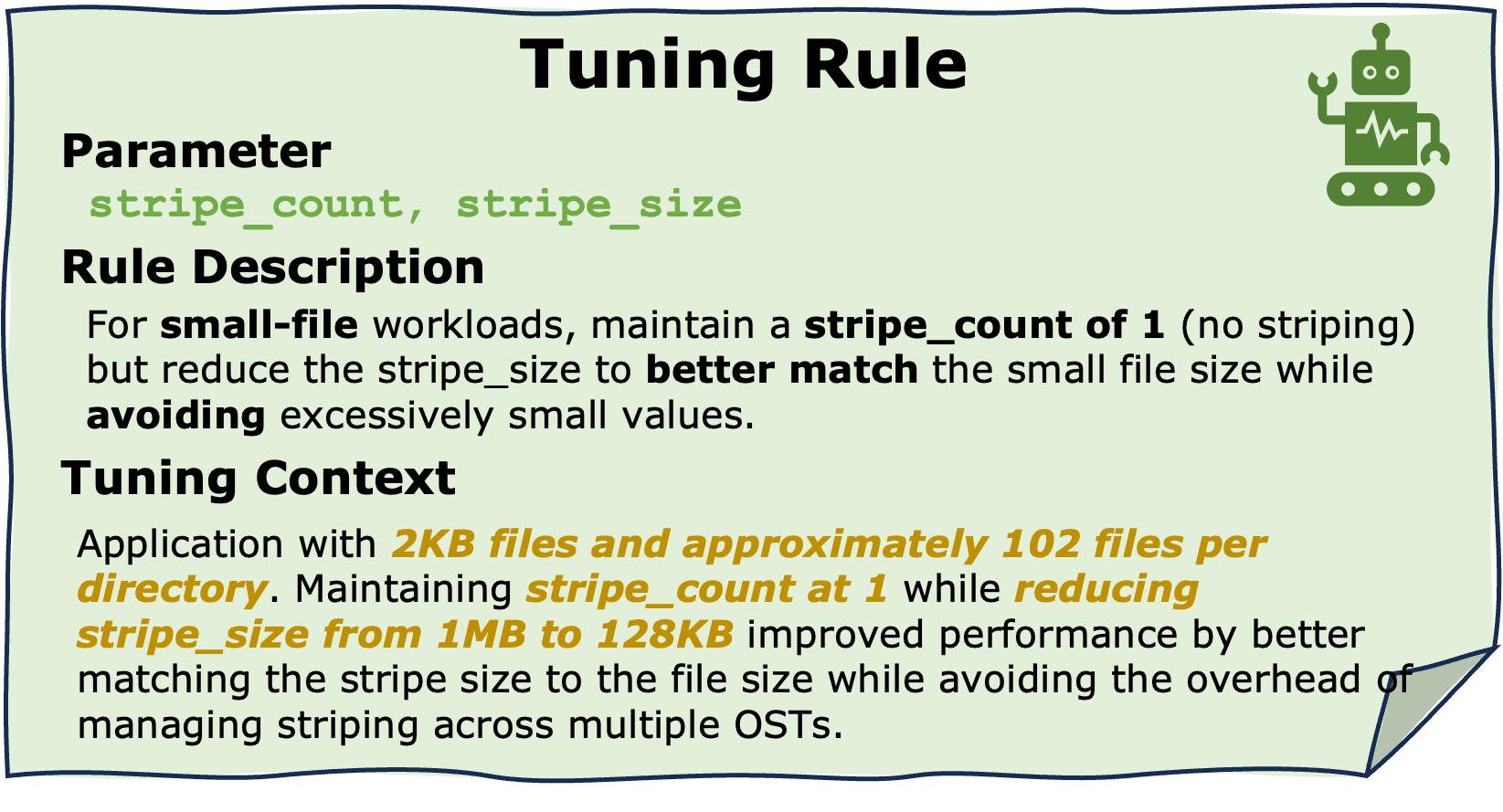}
\vspace{-1em}
    \caption{Example of generated tuning rule.}
        \vspace{-1em}
    \label{rule-sample}
\end{figure}

\subsubsection{Rule Set Design}
Each rule in the rule set generated by the agent refers to one or multiple parameters and contains both a definition for the rule itself and the I/O behavior context in which the rule applies. These are explicitly defined as the \textit{parameter}, \textit{rule description} and \textit{tuning context} in the prompt which instructs the \textit{Tuning Agent} to generate the rule set. Additionally, in the prompt we instruct the \textit{Tuning Agent} to exclude the name of the application being tuned and to make general recommendations as opposed to specific ones. 
%This design is motivated by the intuition that during every procedure, some information should have been learned about tuning one or more individual parameters successfully or unsuccessfully and that this information should not need to be rediscovered in future tuning procedures where similar I/O behavior is found. Additionally, the \textit{tuning context} is added to make easier for the \textit{tuning agent} to know when a given rule should be applied. Finally, the specific instructions to generalize the recommendations in each rule are meant to guide the tuning agent towards the optimal parameter ranges for previously observed I/O behavior while upholding some exploration of the parameter space to avoid local optima.
An example rule generated by the \textit{Tuning Agent} is shown in Figure~\ref{rule-sample}. As illustrated in the rule description, the recommendation for \textit{stripe size} does not specify exact values to try, but instead offers guidance that the setting should be informed by the file size. The \textit{tuning context} also clearly outlines the I/O characteristics of the workload where the rule was learned, making it easier to apply the rule to new workloads with similar characteristics. 
%dong: this reads good to me
\textcolor{black}{To avoid ambiguous rule sets potentially missing important information in each generated rule, we enforce a strict output structure when the LLM generates the rule set. Namely, the LLM must generate a JSON-structured rule set which is organized as a list of objects where each object contains a \textit{Parameter}, \textit{Rule Description}, and \textit{Tuning Context} keys defined with their relevant values.}
%An example of rule generated by the \textit{tuning agent} is shown in figure \ref{rule-sample}. As shown in the rule description, the recommendation for stripe size does not explicitly define values to try but rather provides useful guidance that the setting should be informed by the file size. Also, the \textit{tuning context} clearly defines the I/O characteristics of the workload where the rule was learned which can easily be matched with similar characteristics for new workloads.

\subsubsection{Rule Set Generation and Synthesis}
After the rule set has been generated once, it can serve as a global \textit{Rule Set} and be added to the initial prompt of any subsequent runs, and each subsequent run can serve to update the \textit{Rule Set} to include any new information learned from that run. Concretely, runs following the initial one add the most recent global \textit{Rule Set} to the initial prompt given to the \textit{Tuning Agent}. Once the \textit{Tuning Agent} ends the tuning process, it is asked to augment the existing set of rules rather than generate a completely new set. Since existing rules can conflict with new ones, the rule synthesis process can resolve these conflicts in two ways depending on the nature of the conflict. First, when a new rule directly contradicts an existing rule, the tuning agent is asked to remove it from the rule set. For example, if there exists a rule for the same parameter and equal tuning context as a new rule but the definitions suggest opposite guidance for how the parameter should be tuned, the \textit{Tuning Agent} should remove both of these because it cannot be determined which is more correct. Alternatively, suppose the tuning context and intended parameter between two rules are equal and the rules offer only slightly different guidance. In that case, they should be noted as alternative approaches so that future tuning attempts could potentially try both. 
\textcolor{black}{
If both approaches are attempted in a future tuning run but only one of them produces a positive outcome, the tuning agent drops the negative approach when updating the rule set during this run, the same JSON structure as mentioned in the previous subsection is reused when merging rule sets to maintain a common format.
}

\begin{figure*}[h!]
    \centering
\includegraphics[width=0.9\textwidth]{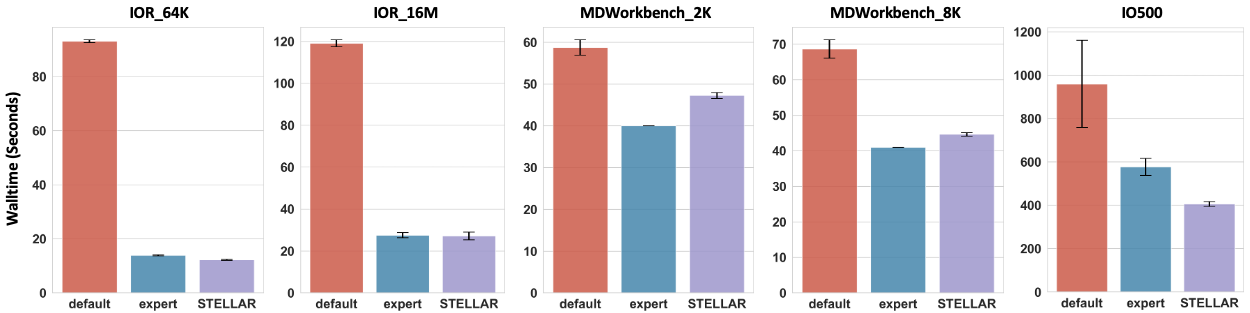}
    \caption{Comparison of \texttt{STELLAR}'s tuning performance with \textit{default} and human \textit{expert} baselines. Smaller values are better.}
    \label{comparison-results}
\end{figure*}

\section{Evaluation}
\label{sec:eval} 
\subsection{Evaluation Setup}
To demonstrate \texttt{STELLAR}’s performance and validate the design decisions outlined earlier, we conducted comprehensive evaluations and present the results in this section. 
%We also include a detailed case study to provide insights into \texttt{STELLAR}’s tuning process. 
Through these evaluations, we aim to answer the following key questions:
\begin{itemize}
\item Within a limited number of attempts, can \texttt{STELLAR} successfully tune the parallel file system for improved I/O performance? If so, how does its performance compare with that of human experts?
\item Can \texttt{STELLAR} effectively accumulate useful knowledge (i.e., a global \textit{Rule Set}) and leverage it to continuously enhance its tuning efficiency?
\item How do individual design choices, such as the RAG-based parameter extraction and the \textit{Analysis Agent}, impact the performance?
%\item How does \texttt{STELLAR} perform tuning for a specific application? What does it observe and how does it reason?
\end{itemize}

We do not include comparison results with traditional machine learning-based autotuners because they all take too long (hundreds of iterations, taking hours or days) to converge to similar results to those \texttt{STELLAR} achieves within five attempts. Also, since they cannot tune 13 parameters at the same time, a side-by-side comparison would be unfair. Instead, We compare with human experts' suggestions when the final I/O performance matters.

%Each of these evaluations were conducted on the same evaluation platform and using the Lustre File System as described in the following section. 

In the following sections we refer to the entire tuning process of \texttt{STELLAR} as a \textit{Tuning Run}, which starts from the application's initial execution to the end of the tuning. Between \textit{Tuning Runs} we always perform the following steps to ensure the results are not contaminated: (1) delete all data files and directories, (2) clear all client-side caches, (3) remount the entire file system on all client nodes, and (4) wait until all queued Lustre sync changes are completed. To ensure that our results are not impacted by noise, we ran each cases eight times to get averages. We show the 90\% confidence interval when needed.

%to a single end-to-end run of STELLAR's online tuning process for a single application from the initial application run under default system settings to the tuning agent's decision to end the tuning process as a \textit{tuning run}. Also, before every execution of a given application before or during a tuning run, these steps are taken to ensure fair application results for every run interpreted by STELLAR: 1) the application's data files and data directories are deleted, 2) all client-side caches are cleared, 3) the file system is remounted on all client nodes, and 4) the execution waits to begin until all queued Lustre sync changes are completed. 

\subsubsection{Evaluation Platform}
All evaluations were conducted on the CloudLab platform~\cite{duplyakin2019design}. CloudLab is an open platform that allows others to easily replicate our experiments. Due to the nature of our work, we were unable to conduct evaluations on large-scale production HPC systems because many of the tunable parameters require root privileges to modify.
Specifically, we allocated 10 CloudLab machines to build a cluster for the evaluations. Each machine is equipped with an Intel Xeon Silver 4114 processor with 10 physical CPU cores and approximately 196 GB of memory and is connected via a 10 Gbps network switch. We installed Lustre 2.15.5, configured with five object storage servers and a combined management server (MGS) and metadata server (MDS)~\cite{braam2019lustre}. The remaining five machines were used as client nodes to run benchmarks and real applications.

%The presented evaluations were conducted on a cluster with 11 linux machines, each running version 2.15.5 of the Lustre~\cite{braam2019lustre} parallel file system, allocated on the Cloudlab platform \cite{duplyakin2019design}. Each machine is equipped with an Intel Xeon Silver 4114 processor with 10 physical cpu cores, roughly 196 GB of memory and linked via a 10 gbps network switch. The Lustre file system is configured with 5 Object Storage Targets (OSTs) on separate machines, 1 management server (MGS) and metadata server (MDS) both configured on a single machine, while the remaining 5 machines are used as client nodes.

\begin{figure*}[h!]
    \centering
\includegraphics[width=1.0\textwidth]{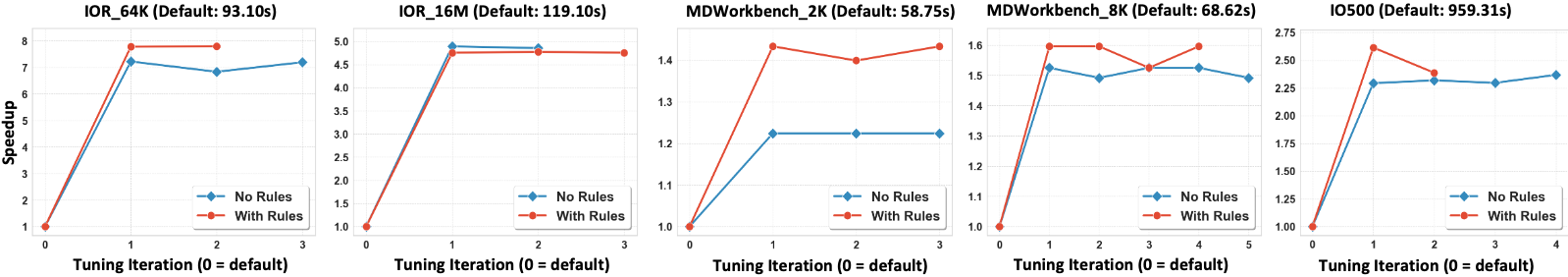}
    \caption{Speedup compared with default Lustre settings with and without the global \textit{Rule Set}. Larger values are better.}
    \label{RS-benchmarks}
\end{figure*}

\subsubsection{Benchmarks}
We selected three classic HPC I/O benchmarks - IOR~\cite{iorMDTest}, MDWorkbench~\cite{kunkel2018understanding}, and IO500~\cite{io500} - to systematically evaluate how \texttt{STELLAR} handles I/O-intensive, metadata-intensive, and mixed workloads. All benchmarks ran in parallel using 50 MPI processes across five client nodes. 

%The primary set of applications used to evaluate STELLAR consists of 5 benchmark applications all of which were run using 50 MPI processes across 5 client nodes. 
We created two workloads using IOR. The first, labeled as \textit{IOR\_64K}, has each MPI process concurrently write/read a 128 MB block using 64 KB transfer size. The I/Os were conducted randomly to a shared file. The second, labeled as \textit{IOR\_16M}, has each MPI process write/read three 128 MB blocks using a large transfer size of 16 MB with a sequential access pattern to a shared file. These two workloads represent two representative I/O patterns: random small and sequential large writes.

%Two of these applications were created with the HPC benchmarking code, IOR\cite{iorMDTest}, which can model intense read/write workloads with configurable characteristics. Our first of these applications, \textit{IOR\_64K}, is set up to have each MPI process write/read a single 128MB block using a transfer size of 64k and a random access pattern to a shared file. The second IOR based applications, \textit{IOR\_16M}, is set up to have each MPI process write/read 3 128MB blocks using a large transfer size of 16MB with a sequential access pattern to a shared file. 
We created another two workloads using MDWorkbench. The first, labeled \textit{MDWorkbench\_2K}, creates 10 directories per process and fills each directory with 400 files, each sized 2 KB. The second, labeled \textit{MDWorkbench\_8K}, also creates 10 directories per process and fills each with 400 files, but each file is 8 KB. Both MDWorkbench workloads ran for three rounds, where each round conducted \texttt{open}, \texttt{write}, \texttt{close}, \texttt{stat}, \texttt{open}, \texttt{read}, \texttt{close}, and \texttt{unlink} operations on each file.
%more of the benchmark applications are created using the MDWorkbench\cite{kunkel2018understanding} benchmarking tool, which is designed to model metadata-intensive workloads which are not easily cached or optimized on distributed file systems. The first MDWorkbench configuration creates 10 directories per process and fills each directory with 400 files, each of size 2KB. The second MDWorkbench configuration also creates 10 directories per process and fills each with 400 files, but each file is 8KB. Both MDWorkbench workloads operate for 3 rounds where each round conducts open, write, close, stat, open, read, close, unlink operations on each file.

The final workload is based on \textit{IO500}~\cite{io500}, which combines IOR and MDTest workloads into one application running through multiple phases including sequential read/write with large access sizes (IOR-Easy), random read/write with small access sizes (IOR-Hard), and metadata-intensive workloads for empty (MDTest-Easy) and small files (MDTest-Hard). This workload challenges autotuners to find the best configurations for the combined workloads.

\subsubsection{Real Applications}
In addition to benchmarks, which were picked for better demonstration, we used a set of real applications to evaluate \texttt{STELLAR}. Specifically, to be representative, we selected one scientific application I/O kernel and one I/O proxy application. The scientific application I/O kernel uses the AMReX framework \cite{zhang2021amrex}, which implements highly concurrent, block-structured adaptive mesh refinement. The I/O proxy application originates from MACSio \cite{miller2015design}, which is designed to model I/O workloads from multiphysics applications primarily, with highly variable data object distribution and composition. Since MACSio's object size and can be configured to take on various sizes, we evaluate one configuration using an object size of 512KB (labeled as \textit{MACSio\_512K}) and another configuration using an object size of 16MB (labeled as \textit{MACSio\_16MB}). Similar to the benchmark applications described previously, our evaluations of these applications were conducted using 50 MPI processes spread across 5 client machines.

\begin{figure*}[h!]
    \centering
\includegraphics[width=0.7\textwidth]{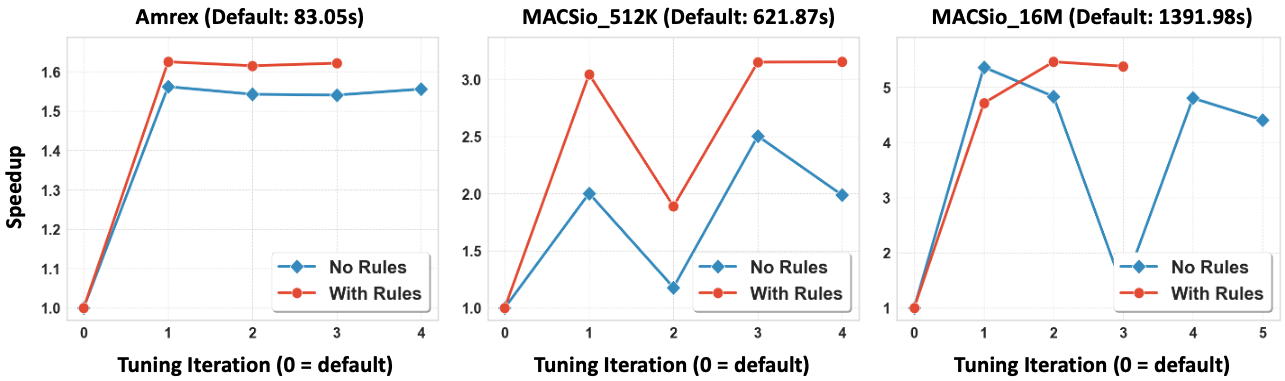}
    \caption{Speedup compared with default Lustre settings with and without a global \textit{Rule Set} for real applications. For all the results, larger values are better.}
    \label{RS-real-apps}
\end{figure*}

\subsection{\texttt{STELLAR} Tuning Performance}
Our first evaluation answers whether \texttt{STELLAR} can successfully tune a PFS for improved I/O performance within a limited number of attempts.

%To achieve this, we compared the final performance of the five benchmark workloads after \texttt{STELLAR}'s tuning. Here, we allowed \texttt{STELLAR} to conduct only five attempts maximally. We compared the results with both default system baseline (\textit{default}) and human expert baseline (\textit{expert}). Note that we compare the end-to-end running time of these benchmarks; hence smaller results are better.

In these evaluations, \texttt{STELLAR} starts from fresh, without any previous knowledge about these workloads nor any global \textit{Rule Set}. The results are reported in Figure~\ref{comparison-results}. Here, \textit{default} refers to the configuration of default parameter settings in Lustre. The \textit{expert} refers to the configuration provided by an I/O expert. To help the expert make better tuning decisions, in this case,  we provided the full information about the benchmarks, including a full description of the benchmark settings and the full Darshan trace logs. The expert was also given practically unbounded time to generate the suggested set of parameters. 
%highlights STELLAR's tuning performance compared to a default system baseline as well as human expert baseline on 5 unique benchmark applications. This is to show how STELLAR is able to converge near or beyond human level performance without any prior knowledge of the applications (i.e. no rule set) in a single tuning run of at most 5 generated configurations.
%To conduct this evaluation, we gathered average performance over 8 runs of three different parameter configurations for each of the previously described benchmark workloads. For each application, the first configuration we ran is the \textit{default} which refers to the configuration representing the default parameter settings of the Lustre file system. 
%The second configuration represents the set of parameter settings suggested by a system expert for each individual application. In this case, a system expert was given the full information of the application being tuned for, including a full description of the benchmark settings used for the run and the I/O trace logs generated by running the application once with the default storage system parameters. They were also given practically unbounded time to generate the suggested set of parameters. We refer to the result of running with this configuration as \textit{expert}. 
The \texttt{STELLAR} results represent the best configuration generated from the tuning run. Note that we limited STELLAR to try at most five configurations and forced it to stop if not automatically stopped by that limit.
%The result of rerunning the selected configuration is referred to as \textit{STELLAR}.
%We present the comparison of results on each of the benchmark applications with each of the three configurations in figure \ref{comparison-results}. 
Each bar represents the wall time (in seconds) of the workload; hence smaller values are better. We show the average  and include a 90\% confidence interval for each average to show the expected performance variance for each configuration. As indicated by these results, \texttt{STELLAR} was able to generate very high-performance parameter configurations that perform much better than the default and perform similarly to or even better than human-level performance. Notably, in the case of IO500, \texttt{STELLAR} was able to outperform the human expert baseline, which proves its ability to adapt to workloads that have multiple phases and variable I/O patterns.
%\textcolor{blue}{add details of selection comparison to human expert}

\subsection{\texttt{STELLAR} Tuning with Global Rule Set}
In our second evaluation we answer two questions: (1) Can \texttt{STELLAR} successfully and consistently interpolate tuning rules as it tunes different applications? (2) Can the accumulated tuning rules be extrapolated to new and unseen applications?

%highlight the impact which the \textit{Global rule Set} has on three aspects of the tuning process. Namely, the rule set should lead to a high-performance initial guess for optimal parameter settings as rules provide the tuning agent with additional useful context based on historical experience to quickly find value ranges which work well for observed I/O patterns. Additionally, the rule set should stabilize the tuning process by avoiding parameter settings which are known to not work well for specific I/O behaviors. Finally, the rule set should allow the tuning process to be shortened as less exploration is required if the rules apply as expected. To highlight these aspects we evaluated the rule set in two scenarios. 

\subsubsection{Rule Set Interpolation}
In the first scenario we used \texttt{STELLAR} to tune all the benchmark applications one by one without any rule set first. Since each tuning run will create some tuning rules, \texttt{STELLAR} needs to merge them into the global \textit{Rule Set}. It is critical that such merging operations do not generate wrong or inconsistent rules. To verify that, we again tuned all the benchmark applications, but this time with the global \textit{Rule Set} applied. 

%This scenario evaluates STELLAR's ability to interpolate the knowledge from a rule set since the workloads being tuned with the rule set are represented in the rule set. 
The results of this evaluation are presented in Figure \ref{RS-benchmarks}, where each plot corresponds to a single benchmark application. Each plot's $x$-axis indicates the number of tuning iterations that STELLAR conducted (i.e., number of configurations tried) until the \textit{Tuning Agent} decided to end the process. Note that iteration 0 indicates the initial run without any tuning. The $y$-axis indicates the speedup compared with the average performance of eight runs using the default parameter settings. 
The results strongly support the fact that \texttt{STELLAR} successfully interpolates knowledge from the tuning rules. Four of the five cases show that leveraging the generated global \textit{Rule Set} resulted in a significantly improved first guess at the optimal parameter settings. In fact, in the case of \textit{MDWorkbench\_2K}, where our previous results (i.e., Figure~\ref{comparison-results}) showed the human expert-generated configuration to be slightly better, adding the rule set helped \texttt{STELLAR} fill this gap and generate configurations that perform equally well as the expert-generated configuration. Further, because of the high performance of the initial guess, \texttt{STELLAR} required less additional exploration of the parameter settings, leading to faster conclusion of the tuning process in 3 out of 5 cases and equal amounts in 1 out of 5. Again, these results further confirmed that \texttt{STELLAR} tunes the PFS within five attempts.

\subsubsection{Rule Set Extrapolation}
In the second scenario we first used \texttt{STELLAR} to tune each of the real application workloads without a rule set and then tune these applications again using the rule set generated accumulated from only tuning the benchmark workloads. 

In contrast to the previous scenario, this scenario evaluates \texttt{STELLAR}'s ability to extrapolate knowledge from the learned rule set to previously unseen workloads.
The results are shown in Figure \ref{RS-real-apps}, where each plot corresponds to a single real application. We can easily observe, even in such a more difficult scenario, that the positive impacts of the global \textit{Rule Set} hold. This is evidenced by the fact that in all cases, the rule set enabled more stable convergence and higher performance of each generated configuration on average. Notably, for the \textit{MACSio\_16M} application, the rule set helped avoid exploration of configurations that had similar performance to the default settings, which were explored without the rule set. Similarly, for \textit{MACSio\_512K}, the addition of the rule set helped avoid the worst settings that were explored without the rule set.

\begin{figure}[t]
    \centering
\includegraphics[width=0.7\columnwidth]{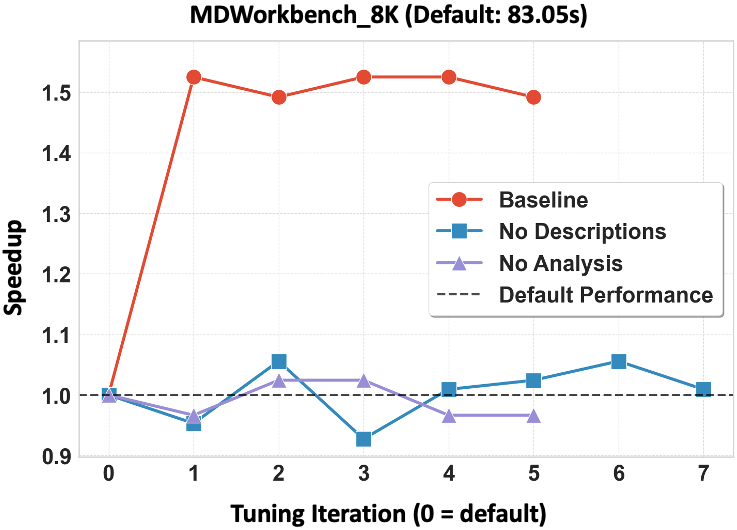}
    \caption{Tuning performance on MDWorkbench with removed STELLAR components.}
    \label{ablation}
\end{figure}

\subsection{\texttt{STELLAR} Ablation Tests}
This evaluation answers the third question: how each component of \texttt{STELLAR} contributes.
We conducted two ablation tests to validate STELLAR's primary components. Because of space limitations, we show only the results conducted on the \textit{MDWorkbench\_8K} benchmark workload since previous results showed that \texttt{STELLAR} required more iterations for this workload, indicating it is more difficult to tune. 

In our first ablation test we evaluated the impact of the RAG-based Parameter Extractor. Specifically, we removed the parameter descriptions generated via the RAG process and observed the tune process of \texttt{STELLAR}. The results are shown in Figure \ref{ablation}, labeled as \textit{No Descriptions}. 
Note that for this test we maintained the valid value ranges for each parameter, which were also generated via our RAG process. This was required because missing these value ranges causes the tuning to fail in most cases when the \textit{Tuning Agent} would often attempt to set invalid values. 
From the results, we can observe a significant drop in performance when parameter descriptions are missing. Upon further analysis of the tuning agent's decisions during this run, we concluded that the primary reason for this drop is the \textit{Tuning Agent}'s lack of accurate understanding of the parameters. For example, when generating \textit{stripe settings}, it understands that a stripe count of 1 is more efficient for small files but then states that changing the parent directory's stripe count to -1 could "distribute the files more evenly across all OSTs." This is a flawed interpretation of how stripe count affects the files in a directory. This misunderstanding can be avoided by our RAG-based extractor, which defines the \textit{stripe count} parameter clearly as "the number of Object Storage Targets (OSTs) across which \textit{a file} will be striped". Avoiding such misinterpretations is critical to avoiding misguided tuning decisions and failing to converge to better parameter settings.
%when the parameter descriptions are present as the definition for the 

%one of the tuning agent's tools, which is to ask the analysis agent for 
Our second ablation test evaluates the impact of removing the \textit{Analysis Agent} from the tuning process. As described in preceding sections, the \textit{Analysis Agent} is a key component in the tuning process as it is tasked with conducting accurate analysis of applications' I/O behavior, which helps guide the \textit{Tuning Agent} toward reasonable initial predictions. Removing the \textit{Analysis Agent} removes both the initial I/O report and the ability to answer additional clarifications from the \textit{Tuning Agent}. The result of removing these features is also catastrophic, as presented in Figure \ref{ablation}, labeled \textit{No Analysis}. The performance is similar to the previous as the \textit{Tuning Agent} fails to generate configurations that significantly outperform the default settings. Additionally, the reason for this level of degraded performance is clearly indicated by the \textit{Tuning Agent}'s decisions, since in this case the agent attempts to increase readahead and RPC size-related parameters. Detailed information on the application's I/O behavior, especially its predominant access to many very small files, would have ruled out such misguided parameter settings.

\begin{figure}[t]
    \centering
\includegraphics[width=0.7\columnwidth]{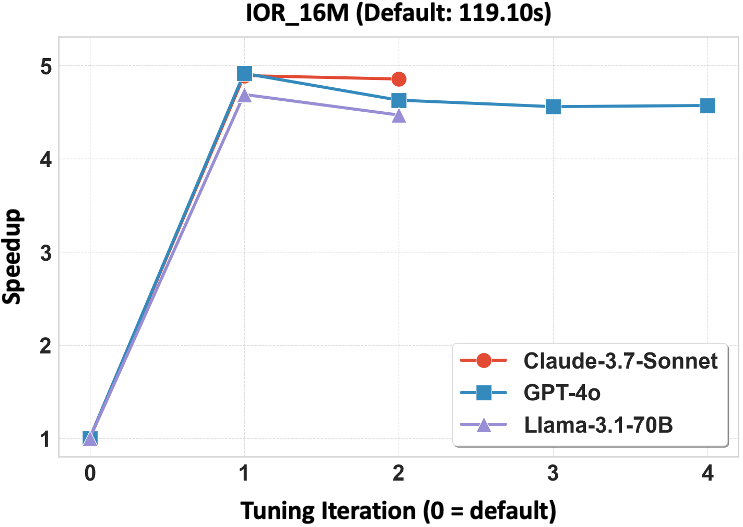}
    \caption{Tuning performance on \textit{IOR\_16M} with different LLMs as the \textit{Tuning Agent}.}
    \vspace{-1em}
    \label{model-comparison}
\end{figure}

\subsection{Model Comparison}
While each of the preceding evaluations was conducted with Claude-3.7-Sonnet acting as the \textit{Tuning Agent}, \texttt{STELLAR}'s performance is not determined by this choice. In fact, any tool-calling LLM should be able to act as \texttt{STELLAR}'s tuning agent. To showcase this, we conducted tuning runs with up to five iterations each for the \textit{IOR\_large} benchmark application with two different LLMs acting as the \textit{Tuning Agent}: OpenAI's GPT-4o and Meta's significantly smaller open source Llama-3.1-70B-Instruct model \cite{grattafiori2024llama3herdmodels}. The results are reported in Figure \ref{model-comparison} and clearly highlight the fact that all three models are able to generate configurations that perform similarly and achieve significant speedups (up to x4.91) compared with the default configuration.

\subsection{Scalability Limitation and Future Directions}
As discussed, all our experiments were conducted on CloudLab due to root-access requirements for system-level parameter tuning. While this constraint limited our evaluation scale, we discuss here the implications for production HPC systems and our path toward broader deployment in the future.

We accept that large-scale systems expand the configuration space and may introduce scale-dependent phenomena. For instance, systems with thousands of nodes enable broader parallelism configurations (e.g., Lustre stripe count and stripe size settings can span wider ranges), while heterogeneous hardware may shift optimal parameter values (e.g., the presence of burst buffers, NVMe-based storage tiers, or varying network topologies can influence the effectiveness of specific parameter combinations). However, we argue that \texttt{STELLAR}'s fundamental approach remains scale-invariant: it automates the same iterative process experts use, regardless of system size, by analyzing the execution, adjusting parameters, and refining based on observations. In fact, larger systems may even facilitate automated tuning by exhibiting more pronounced performance responses to parameter changes, helping \texttt{STELLAR} identify causal relationships more clearly.

Recognizing root-access constraints in production environments, our future work will target user-accessible tuning opportunities: (1) application-layer parameters including HDF5 settings and MPI-IO hints, (2) user-space storage systems like DAOS that provide extensive tunability without privileges, and (3) hybrid approaches where \texttt{STELLAR} recommends both user-controllable and system-level parameters.

\subsection{Cost and Latency Analysis}
%The computational overhead of \texttt{STELLAR} consists of LLM inference time and API costs, both of which are minimal compared to application execution time, which is the dominant factor in the tuning process.
\noindent
\textbf{Latency overhead.} LLM inference introduces only a few seconds of latency per tuning decision across all evaluated models (GPT-4o via OpenAI API, Claude-3-Sonnet via Anthropic API, and Llama-3.1-70B via TogetherAI API). This overhead is negligible compared to application runtime, which typically ranges from minutes to hours for HPC workloads. Thus, the end-to-end tuning time is dominated by application execution rather than LLM processing.

\noindent
\textbf{Token usage and costs.} 
Each of \texttt{STELLAR}'s tuning runs depends on the LLM models behind both the \textit{Tuning Agent} and \textit{Analysis Agent}. For reference, when using Claude-3.7-Sonnet as the \textit{Tuning Agent}, a single complete tuning run processes $\sim$100k input tokens and generates $\sim$13k output tokens on average. Using GPT-4o as the \textit{Analysis Agent}, an equivalent run processes $\sim$400k input tokens and generates $\sim$8k output tokens on average. The total cost of these token generations is highly dynamic and continues to decrease as more performant models become available at lower inference costs. Notably, the nature of the \textit{Tuning Agent's} iterative tuning process, as well as the \textit{Analysis Agent's} analysis process, allows for the majority of key-value matrices calculated for input tokens in each inference request to be cached and reused. In fact, when prompt caching is enabled, between 85 and 90 percent of the total input tokens are resolved via cache over the course of a tuning run. This significantly reduces API costs when using proprietary inference services and reduces computation costs when using local inference platforms.

%In a complete tuning procedure, the Analysis Agent (GPT-4o) processes approximately 400k input tokens and generates 8k output tokens, costing $0.56 for input and $0.15 for output tokens. The Tuning Agent (Claude-3-Sonnet) processes approximately 100k input tokens and generates 13k output tokens, costing $0.30 for input and $0.20 for output tokens. The total cost per tuning run is therefore under $1.50 using commercial APIs.

\begin{figure*}[h!]
    \centering
\includegraphics[width=0.9\textwidth]{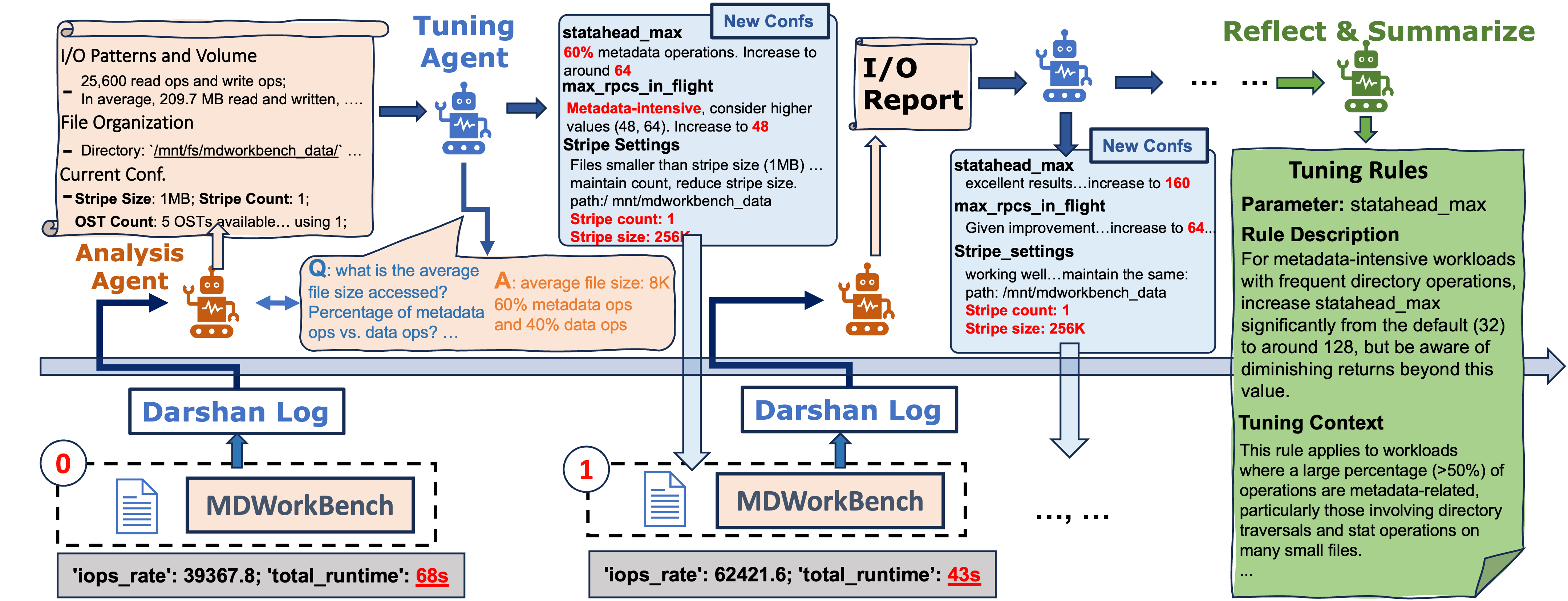}
    \caption{Detailed Tuning Example}
    \label{case-study}
    \vspace{-1em}
\end{figure*}

\subsection{Case Study}
To provide granular insights to the tuning agent's behavior during a tuning run, we present a real case study, shown in figure \ref{case-study}, where STELLAR was tasked with tuning the \textit{MDWorkbench\_8K} benchmark workload. The diagram represents a granular view of the tuning run where initially the application is run using the default parameter settings and the generated I/O trace is analyzed by the Analysis Agent to create the initial I/O report (top left). The Tuning agent is able to effectively analyze the report to ask for useful follow up analysis such as more detailed file size information and metadata to data operation ratios. Using the information from the original report and answers to follow up questions to classify the application's I/O as heavily metadata intensive, the tuning agent can make a high quality first prediction leading to a x1.58 speedup. Based on the positive result, the tuning agent reasonably explores more aggressive changes in the same direction. While the later application results are not shown due to space limitations, the tuning agent found that further changes led to diminishing returns for most parameters. After two rounds of such exploration, the tuning agent finalized the tuning process and summarized what it learned by generating a rule set. One of the generated rules is presented at the end of the timeline and shows the the tuning agent is effectively able to summarize its interactions over the entire tuning process to extract the most important information which it learned while tuning the parameter. Additionally, the details provided by the analysis agent early in the tuning run, allow the tuning agent to assign a detailed context in which the rule was learned so that it cn be applied in similar contexts in future runs.

\section{Conclusion and Future Plan}
\label{sec:conclude}
In this study, we proposed \texttt{STELLAR}, an autonomous tuner for high-performance parallel file systems. By leveraging agentic LLMs, \texttt{STELLAR} consistently selects near-optimal configurations within the first five attempts, demonstrating a level of efficiency that closely mirrors human expertise. This human-like capability sets \texttt{STELLAR} apart from traditional autotuning methods, which often require hundreds or even thousands of iterations to converge.
%Several key design elements contribute to \texttt{STELLAR}’s effectiveness, including accurate RAG-based parameter extraction, external tool integration, LLM-powered reasoning, and a robust multi-agent architecture. Our comprehensive evaluation results affirm the advantage of these design choices.
We believe \texttt{STELLAR} opens a promising new direction for autonomously optimizing complex HPC infrastructure. Beyond its research value, its practical implementation can democratize I/O performance tuning for domain scientists, ultimately accelerating scientific discovery.
In the future, we plan to extend \texttt{STELLAR}’s evaluation to larger-scale systems, potentially in production systems, with the focus on user-accessible tuning parameters.

\section*{Acknowledgements}
We sincerely thank the anonymous reviewers for their valuable feedback. This work was supported in part by the National Science Foundation (NSF) under grants CNS-2008265 and CCF-2412345. This effort was also supported in part by the U.S. Department of Energy (DOE), Office of Science, Office of Advanced Scientific Computing Research (ASCR) under contract number AC02-06CH11357 with Argonne National Laboratory.
%Argonne National Laboratory's work is supported by the U.S. Department of Energy, Office of Science under DE-AC02-06CH11357.

\begin{comment}
\clearpage
\noindent\fbox{\begin{minipage}{\columnwidth}
The submitted manuscript has been created by UChicago Argonne, LLC, Operator
of Argonne National Laboratory ("Argonne”). Argonne, a U.S. Department of
Energy Office of Science laboratory, is operated under Contract No.
DE-AC02-06CH11357. The U.S. Government retains for itself, and others acting
on its behalf, a paid-up nonexclusive, irrevocable worldwide license in said
article to reproduce, prepare derivative works, distribute copies to the
public, and perform publicly and display publicly, by or on behalf of the
Government. The Department of Energy will provide public access to these
results of federally sponsored research in accordance with the DOE Public
Access Plan (http://energy.gov/downloads/doe-public-access-plan).
\end{minipage}}
\end{comment}

\bibliographystyle{plain}
\bibliography{references}

\end{document}